\documentclass[12pt]{iopart}
\usepackage[dvips]{graphicx, color} 
\begin{document}

\title{Magnetic and  electric properties of CaMn$_7$O$_{12}$ based multiferroic compounds: effect of electron doping}
\author{J. Sannigrahi$^{a}$, S. Chattopadhyay$^{a}$, D. Dutta$^{b}$, S. Giri$^{a}$, S. Majumdar$^{a*}$}
\address{Department of Solid State Physics, Indian Association for the Cultivation of Science, 2A \& B Raja S. C. Mullick Road, Jadavpur, Kolkata 700 032, India}
\address{$^{b}$Department of Metallurgical Engineering \& Materials Science, Indian Institute of Technology Bombay, Powai, Mumbai 400 076} 
\ead{$^{*}$sspsm2@iacs.res.in}
\markboth{{\it{\small Magnetic and  electric properties of CaMn$_7$O$_{12}$..... }}}{{\it{\small Magnetic and  electric properties of CaMn$_7$O$_{12}$..... }}}

\begin{abstract}

The mixed-valent multiferroic compound CaMn$_7$O$_{12}$ is studied for its magnetic and electric properties. The compound undergoes  magnetic ordering below 90 K with a helimagnetic structure followed by a low temperature  magnetic anomaly  observed around 43 K.  The present study shows that the magnetic anomaly at 43 K is associated with  thermal hysteresis indicating first order nature of the  transition. The compound also shows field-cooled magnetic memory and relaxation below 43 K, although no zero-field-cooled memory is present. Clear magnetic hysteresis loop is present in the magnetization versus field measurements signifying the presence of some ferromagnetic clusters in the system. We doped trivalent La at the cite of divalent Ca expecting to enhance the fraction of Mn$^{3+}$ ions. The La doped samples show reduced magnetization, although the temperatures associated with the magnetic anomalies remain almost unaltered. Interestingly, the spontaneous electrical polarization below 90 K increases drastically on La substitution. We propose that the ground state of the pure as well as the La doped compositions contain isolated superparamagnetic like  clusters, which can give rise to metastability in the form of field-cooled memory and relaxation. The ground state is not certainly spin glass type as it is evident from the absence of zero-field-cooled memory and frequency shift in the ac suceptibility measurements. 
\end{abstract} 
\pacs {75.85.+t, 75.60.Ej, 75.30.Kz}
\maketitle

\section{Introduction}
In recent times there have been tremendous interest in  multiferroic materials for their versatile technological importance~\cite{mathur, cheong,fiebig, lawes1}. A multiferroic possesses both ferroelectricity and magnetic order, and it is particularly important when the magnetic and electric phenomena remain intercoupled. Such magnetoelectric (ME) effect is quite promising for memory devices, and at the same time has fundamental importance. 

\par
Ferroelectricity and magnetism in a material are generally mutually exclusive and even when they occur together in a material they have their different sources. As a result the ME coupling is found to be quite weak. Examples of such multiferroics are found among transition metal oxides such BiFeO$_3$, BiMnO$_3$ etc~\cite{catalan,kimura}. However, in last one decade several other transition metal oxide based materials were discovered where the magnetic order itself is responsible for the development of ferroelectric polarization~\cite{khomskii}. Such materials are often referred as magnetic multiferroic and examples include rare-earth  manganites (TbMnO$_3$, DyMnO$_3$,YMnO$_3$,  TbMn$_2$O$_5$ etc.)~\cite{goto, hur}, vanadates (Ni$_3$V$_2$O$_8$)~\cite{lawes2}, chromate (CoCr$_2$O$_4$) etc. The existence of direct coupling  between magnetic and electric polarizations in these material are of fundamental importance. However, ME coupling strength is still rather weak in these materials and the  value of spontaneous electric polarization ($P$) is few order of magnitude smaller than the conventional proper ferroelectrics along with the low value of  ferrolectric transition temperatures. Consequently they are not useful for practical applications.

\par      
Very recently, a new magnetic multiferroic CaMn$_7$O$_{12}$ has been reported in the literature which show large ferroelectric polarization (2870 $\mu$Cm$^{-2}$) below the helical magnetic ordering temperature of 90 K~\cite{zhang,rdj}. CaMn$_7$O$_{12}$ belongs to the quadruple (AA$^{\prime})_3$B$_4$O$_{12}$ family of manganites and has a perovskite derived crystal structure with trigonal space group R$\bar{3}$~\cite{bochu, ws}. In this structure, Mn has three crystallographically inequivalent sites with Wykoff positions 9e, 9d and 3b. The sites 9e and 9d are occupied by Mn$^{3+}$ ions (three Mn-ions at each site per formula unit), and the 3b site is occupied by one Mn$^{4+}$ ion~\cite{zhang,rdj,rp1,rp2}. As evident, the compound is mixed valent with the Mn$^{3+}$ and Mn$^{4+}$ ratio being 6:1. The Mn$^{3+}$ (9d) and Mn$^{4+}$ (3b) remain in a charge ordered state below 250 K~\cite{rp2}.

\par 
The compound undergoes two magnetic transitions, one at $T_1$ = 90 K and the other at $T_2$ = 45 K~\cite{zhang,rp1}. Below 90 K, the magnetic structure is helical with incommensurate magnetic propagation vector ${\bf q_1}$ = (0, 1, 0.963)~\cite{rdj}. The magnetic structure is found to be more complex below $T_2$ and it may be represented by two coexisting  propagation vectors namely, ${\bf q_2^{\pm}}$ = ${\bf q_1} \pm (0, 0 , \delta)$, where $\delta \sim$ 0.08. The sample shows giant ferroelectric polarization along $c$ axis below 90 K~\cite{zhang,rdj}. The magnetic nature of such electrical polarization is also confirmed by the strong ME effect observed below 90 K, which shows about 30\% change in $P$ under an applied magnetic field of $H$ = 90 kOe~\cite{zhang,ms}. Based on the recent density functional theory (DFT) calculation, it has been argued that giant $P$ and ME in CaMn$_7$O$_{12}$ arise from the symmetric exchange-striction in presence of strong Dzyaloshinskii-Moriya (DM) type magnetic interaction between certain Mn$^{3+}$ and Mn$^{4+}$ ions~\cite{lu}.  

\par
It is understood that the magnetic and ferroelectric anomalies in CaMn$_7$O$_{12}$ are strongly governed by the  mixed valent nature of manganese. In the present work we carefully examined the nature of the magnetic transitions as well as their evolution with changing Mn$^{3+}$ and Mn$^{4+}$ ratio. We substituted trivalent La at the divalent Ca to tune the Mn$^{3+}$/Mn$^{4+}$ ratio. Our study indicate interesting magneto-thermal irreversibility across the low temperature magnetic transition along with drastic change in the magnetic and ferroelectric properties. 
     
\begin{figure}[t]
\centering
\includegraphics[width = 8 cm]{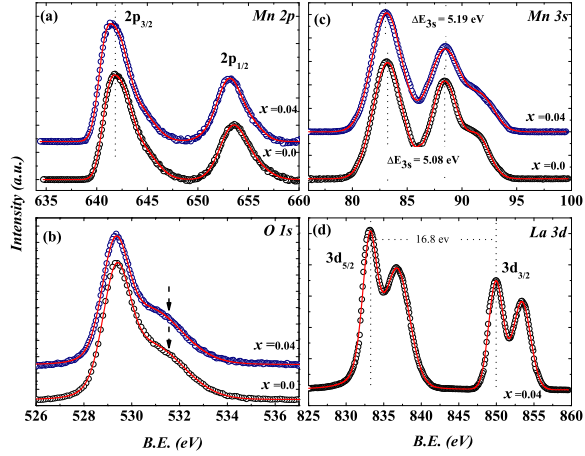}
\caption {Core level x-ray photoelectron spectra of (a) Mn-2$p$, (b)O-1$s$, (c) Mn-3$s$, and (d)La-3$d$ levels recorded at room temperature for CaMn$_7$O$_{12}$ and Ca$_{0.96}$La$_{0.04}$Mn$_7$O$_{12}$ samples.}
\label{xps}
\end{figure}

\section{Experimental Details}
Ceramic  samples of  CaMn$_7$O$_{12}$ and some La doped compositions (Ca$_{1-x}$La$_x$Mn$_7$O$_{12}$, $x$ = 0.04, and 0.08 ) were prepared by sol-gel method using polyethylene glycol gel. The gel produced by the method was first heated at 250$^{\circ}$ C in the oven. The products  were  heated consecutively  at (i) 800 $^{\circ}$ C, (ii) 925$^{\circ}$ C, and  (iii) 950$^{\circ}$ C each for 16 h. Finally the powder samples were pelletized and heated 970$^{\circ}$ C for 60 h. Room temperature powder x-ray diffraction (XRD) data of the samples were recorded  using Cu K$_{\alpha}$ radiation. Energy dispersive spectrometry (EDS) on the samples were performed in an FEI Quanta 200 scanning electron microscope. The pure and $x$ = 0.04 samples were also investigated through core level x-ray photo-electron spectroscopy (XPS) at room temperature using Al $K_{\alpha}$ radiation on a laboratory based commercial instrument (Omicron). The magnetic measurements were performed on a vibrating sample magnetometer from Cryogenic Ltd. UK, as well as on a Quantum Design SQUID magnetometer. The pyroelectric current of the sample was measured using a Kithley electrometer (model 6517B) in a Helium closed cycle refrigerator. The ac dielectric measurements were performed using an Agilent E4980A precision LCR meter in the temperature range 10-300 K.

\section{Sample Characterization}
The XRD patterns (not shown here) indicate that all the samples crystallize in trigonal $R\overline{3}$ crystal structure. No extra unindexable reflection was found in XRD patterns indicating that the samples are all single phase in nature. We have determined the lattice parameters of the samples considering a hexagonal unit cell as described in table~\ref{values}. The hexagonal lattice parameters $a_h$ and $c_h$ decrease systematically with La doping.

\par
We performed careful EDS measurement  to find out the elemental ratio of the studied samples. The EDS shows homogeneous concentration of the constituent elements within the accuracy of the method. We carried out  EDS at 8-10 different regions of the samples with effective scanning area of 0.34 $\times$ 0.34 mm$^2$. The mean atomic ratio (La+Ca):Mn was found to be 1:7.05 ($\pm$ 0.067) and 1:6.94 ($\pm$ 0.09) in $x$ = 0.04 and 0.08 samples respectively. The error values indicated in the bracket are the standard deviation of the elemental ratio measured at different parts of a particular sample. The (La+Ca):Mn values are quite close to the expected stoichiometric value of 1:7. The atomic ratio Ca:La was found to be 1:0.05 (expected 1:0.04) and 1:0.088 (expected 1:0.08) in $x$ = 0.04 and 0.08 samples respectively. These values are quite close to the actual stoichiometry considering the fact that the percentage of La in the doped samples is small. 

\par
We performed core level XPS measurement on $x$ = 0.0 and 0.04 samples at room temperature as shows in figs.~\ref{xps}. The background subtraction and peak fitting were performed using the software {\small XPS-PEAK 4.1}. Fig.~\ref{xps} (a) shows the Mn-2$p$ levels of both the samples. Clear splitting is observed in Mn-2$p$ peak which corresponds to 2$p_{\frac{3}{2}}$ and 2$p_{\frac{1}{2}}$ due to spin orbit coupling. The spin orbit splitting in 2$p$ level is found to be 11.7 eV for both the samples. The Mn-2$p_{\frac{3}{2}}$ is broad and asymmetric in the higher binding energy (BE) side. Such broadness and asymmetry have often been ascribed to the mixed valency of Mn in the sample~\cite{taguchi}. Each 2$p$ doublet was fitted using three components, namely Mn$^{3+}$, Mn$^{4+}$ and a satellite peak. The Mn-2$p_{\frac{3}{2}}$ peak is slightly shifted toward lower BE side in $x$ = 0.04 sample. We have looked at the  energy difference between O-1$s$ (see fig. ~\ref{xps} (b)) and Mn-2$p_{\frac{3}{2}}$ levels and it is found to be $\Delta E_{2p-1s}$ = 112.4 and 112.0 eV for $x$ = 0 and 0.04 samples respectively. The lower value of $\Delta E_{2p-1s}$ in $x$ = 0.04 may indicate the larger Mn$^{3+}$ fraction in the La-doped sample as compared to the undoped one~\cite{decorse}.

\par
Fig.~\ref{xps}(c) shows XPS spectra of the exchange splitted Mn-3$s$ level for $x$ = 0.0 and 0.04 samples. The extend of splitting ($\Delta E_{3s}$) depends on the valence state of Mn ions~\cite{vrg} and Mn valence can be expressed by the empirical relation: $v_{Mn} = 9.67 -1.2\Delta E_{3s}$~\cite{beyreuther}. We find $v_{Mn}$ to be 3.22 and 3.08 respectively for $x$ = 0.0 and 0.04 samples. It is to be noted that average Mn valence of CaMn$_7$O$_{12}$ is expected to be 3.14 (considering 6:1 ratio of the Mn$^{3+}$ and  Mn$^{4+}$). The investigation indicates the enhancement of Mn$^{3+}$ ions in the material on La substitution.  

\par
The presence of La in $x$ = 0.04 sample is confirmed by the observation of La-3$d$ doublet peaks at energies 855.3 eV (3$d_{\frac{3}{2}}$) and 838.5 eV (3$d_{\frac{5}{2}}$). Each doublet is further split into two components, which occurs due to the charge transfer from the ligand 2$p$ level to the La 4$f$ level~\cite{park}. The spin-orbit splitting between 3$d_{\frac{3}{2}}$ and 3$d_{\frac{5}{2}}$ lines are found to be close to 16.8 eV, which is expected for the La$^{3+}$ state.    

\begin{figure}[t]
\centering
\includegraphics[width = 8 cm]{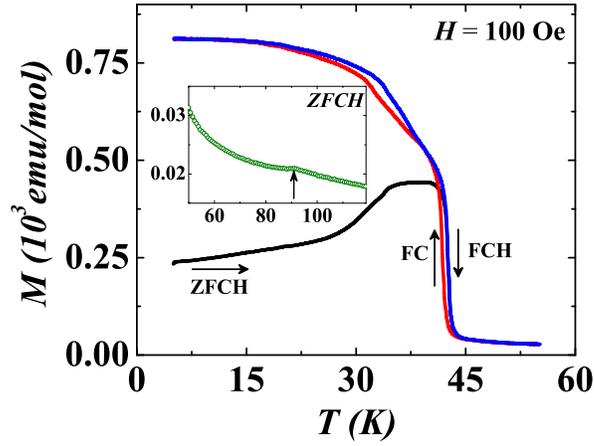}
\caption {Temperature dependence of zero-field-cooled-heating (ZFCH), field-cooling (FC) and field-cooled-heating (FCH) magnetization data of CaMn$_7$O$_{12}$ sample measured in presence of 100 Oe of field. The inset shows the enlarged view of the anomaly at 90 K.}
\label{mt}
\end{figure}

\begin{table*}[t]
\begin{center}
\caption{Variation of hexagonal lattice parameters ($a_h$ and $c_h$), pseudo-cubic angle($\alpha_{cub}$), effective paramagnetic moment($\mu_{eff}$), magnetization in 50 kOe at 5 K $M_{50 kOe}^{5 K}$), coercive field at 5 K ($H_C^{5 K}$), and spontaneous polarization at 10 K ($P^{10 K}$) of Ca$_{1-x}$La$_x$Mn$_7$O$_{12}$ for different values of $x$.}
\begin{tabular}{|c|c|c|c|c|c|c|c|}
\hline
\hline
$x$ & $a_h$ &  $c_h$ & $\alpha_{cub}$& $\mu_{eff}$&$M_{50 kOe}^{5 K}$  & $H_C^{5 K}$ & $P^{10 K}$\\ 
 & (\AA) &  (\AA) & ($^{\circ}$)& ($\mu_B$/Mn)&($\mu_B$/f.u.)  & (kOe)& ($\mu$Cm$^{-2}$) \\
\hline
0.0 & 10.63    &  6.31 & 91.18&4.69 &2.99 & 3.1 & 440\\
0.04 & 10.57    & 6.29  & 91.05&4.78 &2.18  & 0.92& 878 \\
0.08 & 10.51 & 6.28 & 90.93& 4.81&1.84& 0.52&1190\\ 

\hline
\hline
\end{tabular}
\label{values}
\end{center}
\end{table*}    
\section {Results}
\subsection {Magnetization studies on CaMn$_7$O$_{12}$}
Fig.~\ref{mt} shows the dc magnetization ($M$) as a function of temperature ($T$) measured in zero-field-cooled-heating (ZFCH), field-cooled-heating (FCH) and field-cooling (FC) protocols. The ZFCH and FCH data diverges from below about $T_2$ = 43 K. The signature of the high-$T$ magnetic transition is also evident from a kink around 90 K as shown in the inset of fig.~\ref{mt}. However, no signature of the emergence of thermomagnetic irreversibility is observed at this transition. The 43 K transition is more prominent and it is associated with the large enhancement of $M$. We also observe the signature of thermal hysteresis just below $T_2$ between FCH and FC data. Interestingly, FC and FCH data show rather unusual behaviour as far as the thermal hysteresis is concerned. On cooling below $T_2$ they join up at around 38 K and then separated out again to form another hysteresis loop between 38 K and 15 K.  In other words, we observe {\it two separate thermal hysteresis loops} below $T_2$. The existence of thermal hysteresis between 43 and 38 K clearly indicate the first order nature of the magnetic transition at $T_2$, which is also supported by the structural anomaly reported around $T_2$ previously~\cite{ws,ms}. Notably, ac  dielectric permittivity data show an upturn below around 40~K~\cite{ms}, which may be connected to the hysteresis observed in our $M(T)$ data between 38 K and 15 K.  A clear change in slope was reported in the $P(T)$ data around 45 K followed by  a flat region below 35 K~\cite{zhang}. Such anomalies are also apparent in the $M(T)$ data of our measurements.  The anomalous thermal hysteresis observed in our data may indicate the existence of multiple first order phase transitions in the sample. 

\begin{figure}[t]
\centering
\includegraphics[width = 8 cm]{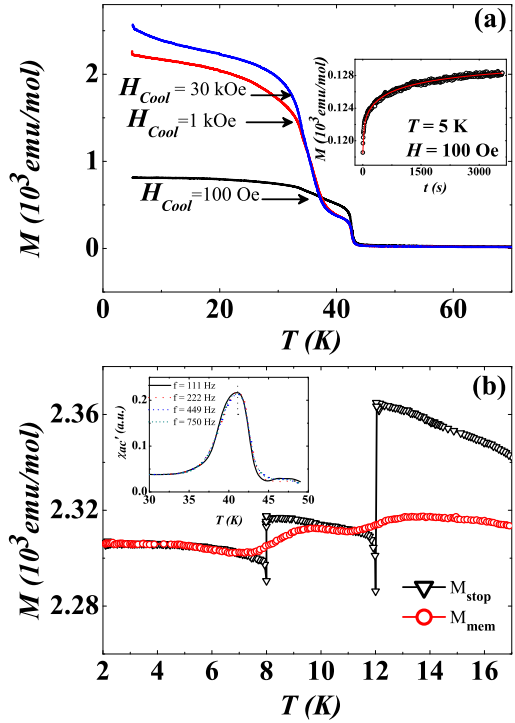}
\caption {(a) Magnetization as a function of temperature measured in 100 Oe while heating after the sample being cooled in three different fields (100 Oe, 1 kOe and 3 kOe). The inset shows the magnetic relaxation curve measured at 5 K. (b) shows the field stop field cooled memory measurement curves. Here $M_{stop}$ was recorded while cooling in 1 kOe with intermediate stops. $M_{mem}$ is the subsequent continuous heating curve measured in 1 kOe. The inset of (b) shows the temperature variation of the real part of ac susceptibility measured at different driving frequencies.}
\label{mem}
\end{figure}

\par
Concomitant occurrence of structural and magnetic anomalies at $T_2$ indicates the  magneto-structural nature of the transition. First order magneto-structural coupling can show intriguing effects as a function of $T$ or $H$. There are examples, where such transition can give rise to metastabilty, arrested dynamics, glassyness etc~\cite{chaddah,roy,chatterjee}. We investigated the thermo-magnetic curves of CaMn$_7$O$_{12}$ measured in different protocols (see fig.~\ref{mem}). We cooled the sample down to 5 K  from 200 K at different applied fields (100 Oe, 1 kOe, and 30 kOe) and measured $M$ while heating in 100 Oe. It is clearly evident that cooling in higher fields produces larger value of $M$.  Field applied during cooling is inducing some magnetic domains/clusters which are oriented favourably along $H$. Such favourably oriented magnetic clusters persist even when the field is lowered and the sample is heated back.
\par
The presence of thermo-magnetic irreversibility and the cooling field effect instigated us to investigate the possible metastable character below $T_2$. The inset of fig.~\ref{mem}(a) shows the time ($t$) variation of $M$ in the ZFC condition at 5 K. The sample was first cooled in absence of $H$ down to 5 K and subsequently 100 Oe of field was applied and then $M$ was measured as a function of $t$. The sample shows large relaxation, where $M$ changes by more than 8\% in 3600 s. Such large relaxation is generally observed in disordered magnetic systems or in spin glasses. The relaxation data can be well fitted by the Kohlrausch-Williams-Watt (KWW) stretched exponential  model ($\sim \exp{[-(t/\tau)^{ \beta}}]$)~\cite{ito,rvc} where $\tau$ is the characteristic relaxation time and $\beta$ is the shape parameter. Such model was widely used to analyze the data for spin glass and other disordered magnets.~\cite{phillips} The value of $\beta$ was found to be 0.37 for the present CaMn$_7$O$_{12}$ sample.  The exponent $\beta$ in the KWW model  signifies the number of intermediate states through which the system should evolve, and it approaches 1 when the number of such intermediate states diminishes.~\cite{xd} Glassy and disordered magnetic systems are found to show $\beta$ values over a wide range between 0.2 to 0.6 and the value of $\beta$ for the present composition falls well within the range. 

\par
We performed magnetic memory experiment~\cite{salamon, gdcu} to further elucidate the metastable state. For the memory measurement, the sample was first cooled in presence of $H$ = 1 kOe  with intermediate stops of 3600 s each at 12 K and 8 K [curve $M_{stop}$ in fig.~\ref{mem} (b)].  At each stop, $H$ was reduced to zero. After reaching 2 K, the sample was heated back in 1 kOe [curve $M_{mem}$ in fig.~\ref{mem}(b)], and the signature of stops is clearly imprinted in the heating curve at respective temperatures in the form of dips. This prominent signature of memory clearly indicate that the system evolves through multiple metastable states as one cools it below $T_2$.

\begin{figure}[t]
\centering
\includegraphics[width = 8 cm]{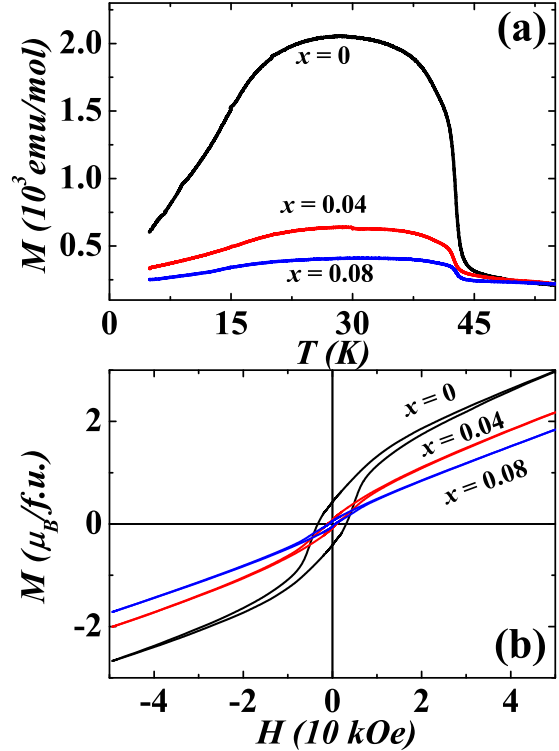}
\caption {(a) Magnetization as a function of temperature in the zero-field-cooled state for Ca$_{1-x}$La$_x$Mn$_7$O$_{12}$ ($x$ = 0.0, 0.04 and 0.08) samples. (b) shows the isothermal magnetization versus field curves at 5 K for different samples.}
\label{mh}
\end{figure}

\par
However, it should be noted that the above memory measurement was performed in the field-cooled condition and positive signature of  memory can be present in assembly of noninteracting super-paramagnetic particles due to their individual relaxation under an applied field~\cite{sasaki,ban}. An elegant way to distinguish a superparamagnet from a glassy magnetic system is the measurement of zero-field-cooled memory. In this protocol, the sample was first cooled in zero field with intermediate stops and subsequently the sample was heated back in presence of small applied field and $M(T)$ data were recorded. For the present sample, although the  field-cooled  memory is very strong, we failed to observe any signature of zero-field-cooled memory (not shown here). We also measured ac susceptibility of the sample around $T_2$ at different driving frequencies ($f$). The signature of magnetic transition at $T_2$ is apparent in the real part of ac susceptibility ($\chi_{ac}^{\prime}(T)$) in the form of a peak. However, we do not see any shift in the peak position with $f$ (see inset of fig.~\ref{mem} (b)). The negative results in zero-field-cooled memory and the absence of $f$ dependency of $\chi_{ac}^{\prime}(T)$ rule out the possibility of a low-$T$ glassy magnetic state of  CaMn$_7$O$_{12}$.           

\begin{figure}[t]
\centering
\includegraphics[width = 8 cm]{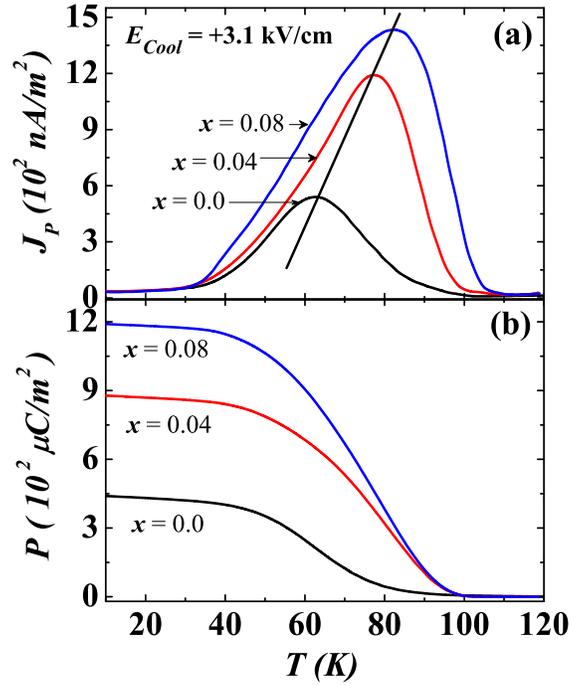}
\caption {(a) and (b) respectively show the temperature dependence of pyroelectric current density ($J_P$) and spontaneous electrical polarization ($P$) for Ca$_{1-x}$La$_x$Mn$_7$O$_{12}$ ($x$ = 0.0, 0.04 and 0.08) samples.}
\label{pt}
\end{figure}

\subsection{Effect of La doping in CaMn$_7$O$_{12}$}
CaMn$_7$O$_{12}$ is a mixed valent compound and it is well known that Mn valency plays a  crucial role in deciding the electronic and magnetic properties of the sample. We tried to vary the Mn valency by doping La at Ca site. Since La remains in the 3+ state as opposed to 2+ state of Ca, it is expected that doping would enhance the Mn$^{3+}$ fraction at the cost of Mn$^{4+}$. Fig.~\ref{mh}(a) shows a comparison of the ZFC $M(T)$ data recorded for Ca$_{1-x}$La$_x$Mn$_7$O$_{12}$ sample with $x$ = 0.0, 0.04 and 0.08. The transition at $T_2$ is present in the doped samples, however they are slightly shifted to lower $T$ with increasing $x$. The most important change due to doping is the drastic reduction of the value of $M$ (see table~\ref{values}). The peak value of $M$ decreases from 3.33 emu/g in pure sample  to 0.65 emu/g for $x$ = 0.08. Such decrease is also  visible in the isothermal $M$ versus $H$ data depicted in fig.~\ref{mh}(b). The coercivity of the magnetic hysteresis curve  gets reduced with increasing $x$. It indicates that the FM-like contribution present in the sample below $T_2$ is getting reduced due to La doping at the Ca site. 

\par
It is to be noted that all the samples ($x$ = 0.0, 0.04 and 0.08)  show paramagnetic behaviour above about 120 K. The Curie-Weiss fitting ($\chi$ = $M/H$ = $C/(T-\theta)$, where $\chi$ is the dc magnetic susceptibility, $C$ = Curie constant and $\theta$ is the Weiss temperature) to the data indicates that the effective paramagnetic moment per Mn ($\mu_{eff}$/Mn) increases with La doping (see table~\ref{values}). This is expected as La doping enhances the Mn$^{3+}$ fraction at the cost of Mn$^{4+}$, and the former spin state has larger value of the isolated magnetic moment (4.9 $\mu_B$) than the later (3.87 $\mu_B$). Using these moment values  along with the calculated $\mu_{eff}$/Mn from Curie-Weiss law, one can estimate the effective valency of Mn in the samples. For $x$ = 0.0, 0.04 and 0.08 samples, the  effective valency of Mn was found to be 3.22, 3.12 and 3.09 respectively. These are  close to the values expected for nominal stoichiometric compositions (3.15, 3.14 and 3.13), and they match well with the $v_{Mn}$ calculated from XPS data.   

\par
It was already shown in the literature that CaMn$_7$O$_{12}$ depicts spontaneous electrical polarization below $T_1$. We have presented here the polarization data as calculated from the measured pyroelectric current density ($J_P$) of the  samples. In order to record $J_P$, we used capacitor type assembly with a pair of electrodes attached to two flat surfaces of the sample using silver epoxy. The sample was first cooled down to 10 K in presence of an electric field, $E_{cool}$ = 3.1 kV/cm. After reaching 10 K, $E_{cool}$ was set to zero and $J_P$  was measured  while the sample was heated at a constant rate of 4 K/min.  One can calculate $P$ by integrating $J_P(t)$. Here we assumed that $P$ vanishes for all the samples as soon as $T$ goes above 120 K~\cite{kohn,kk}. The measured $J_P$ and $P$ have been shown in fig.~\ref{pt} (a) and (b) respectively as a function of $T$. Clearly, a drastic increase in the value of $P$ is observed in the La doped samples(see table~\ref{values}). The peak value of $J_P$ increases by a factor of 2.5 in $x$ = 0.08 sample as compared to the undoped one. Consequently, $P$ is found to be higher in the La-doped compositions. For all the samples, $P$ rises with decreasing $T$ below $T_1$ and eventually saturates at low $T$. The magnitude of this saturated $P$ in our study  is found to be higher in the pure  CaMn$_7$O$_{12}$  ($\sim$ 440 $\mu$Cm$^{-2}$ for $E_{cool}$ = 3.1 kV/cm) compared to the previous report on polycrystalline sample ($\sim$ 240 $\mu$Cm$^{-2}$ for $E_{cool}$ = 3.5 kV/cm)~\cite{zhang}. But the observed value is certainly much smaller than the reported single crystal data ($\sim$ 2870 $\mu$Cm$^{-2}$ for $E_{cool}$ = 4.4 kV/cm)~\cite{rdj}. The important observation is that the low $T$ saturated value of $P$ increases drastically in the La-doped samples and it is about three times higher in $x$ = 0.08 sample.

\begin{figure}[t]
\centering
\includegraphics[width = 8 cm]{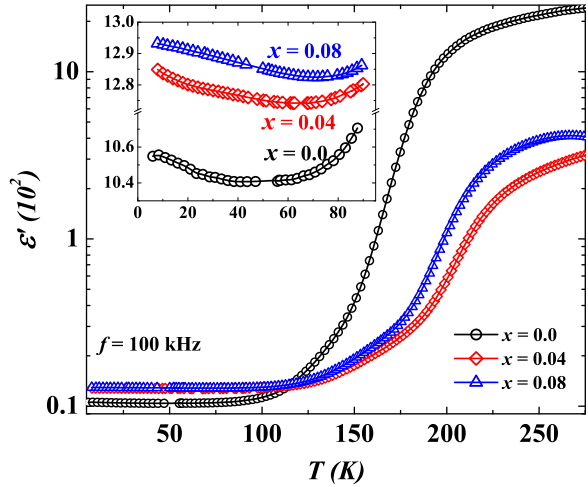}
\caption {Temperature dependence of the real part of complex dielectric permittivity  for Ca$_{1-x}$La$_x$Mn$_7$O$_{12}$ ($x$ = 0.0, 0.04 and 0.08) samples. The inset shows the enlarged view of the low temperature feature.}
\label{ep}
\end{figure}

\par
The electrical response of the samples were further probed by ac dielectric measurements as shown in fig.~\ref{ep}. The real part of the complex permittivity shows $T$-independent flat region at low $T$, followed by a thermally activated rise at high $T$. Such activated behaviour is quite common among ceramics and it is contributed by mobile charge carriers in presence of grain and grain boundary. If we compare the $\epsilon^{\prime}$ with $P$, this flat portion actually corresponds to the region where spontaneous polarization is present. Very similar to $P$, the magnitude of $\epsilon^{\prime}$ in the flat region increases with La doping. Interestingly, the flat region in $\epsilon^{\prime}$ at low-$T$ is not exactly flat, rather it shows a shallow minimum, which was earlier reported for the pure sample~\cite{ms}. It was also found that magneto-capacitive effect is visible around this shallow minimum. For our case, the minimum shifts to higher $T$ with increasing La doping. This is at par with the polarization data, where the temperature corresponding to the appearance of spontaneous polarization  shifts to higher $T$ with $x$.    

\section{Discussion}
The multiferroic compound CaMn$_7$O$_{12}$ is an improper ferroelectric, where spontaneous electrical polarization occurs due to the helical spin ordering below $T_1$ = 90 K~\cite{zhang, rdj, lu}. Our magnetization measurements indicate clear thermal hysteresis around $T_2$, which signifies a first order phase transition (FOPT). Previous x-ray and neutron diffraction  studies~\cite{ws} indicated that the system undergoes structural transition around $T_2$, which is associated with sharp change in lattice parameters. The observation of thermal hysteresis is in line with the previously reported structural anomaly. Although $P$ develops just below $T_1$, it shows a change in slope around $T_2$, which is clearly  associated with the FOPT occurring at this point.

\par
In our magnetic study, we observe interesting magneto-thermal anomaly, which is reflected in the $M(T)$ data measured during heating in 100 Oe after field-cooling in different $H$ through $T_2$. The sample  depicts strong signature of field stop magnetic memory below $T_2$. These observations suggest a metastable ground state of this material.  Recent neutron diffraction data revealed that CaMn$_7$O$_{12}$ has a complex magnetic structure below $T_2$ with a possibility of multiple magnetic propagation vectors and it is difficult to determine the exact magnetic structure without single crystal neutron diffraction data. The rise in $M$  below $T_2$ may be an indication of the formation of superparamagnetic like clusters. On field-cooling, the number of such clusters enhances and and they get oriented  favourably  along $H$. This is the likely origin of enhanced value of $M$ in the heating data if the sample is cooled in higher $H$. The superparamagnetic nature of the sample is also supported by the observation of clear loop in the $M(H)$ data, which is unlikely to occur had the sample been a purely helimagnet.  

\par
Superparamagnetism is generally associated with assembly of noninteracting magnetic nanoparticles, and the field-cooled memory is observed below the blocking temperature due to the N\'eel-type relaxation superparamagntic moment of individual particle~\cite{ban, bedanta}. The present investigation was carried out on bulk ceramic sample where the particle size is too large ($\sim$ 1-10 $\mu$m) to have appreciable N\'eel relaxation. However, evidences for superparamagnetic clusters was reported in several bulk magnetic systems (such as manganites, nickelites etc.)~\cite{banerjee, jia, granado, qin, dagotto}. Therefore, it is possible that nanosized magnetic clusters can form in bulk CaMn$_7$O$_{12}$ in the backdrop of a helimagnetic structure and the relaxation of individual nano-cluster is primarily responsible for the observed field-cooled memory. The complete absence of zero-field-cooled memory rules out the possibility of spin-glass or cluster-glass like ground state indicating the lack of substantial spin-frustration among different magnetic specimens. Considering the fact that metastability occurs  below $T_2$ only, the formation of spin clusters can be strongly associated with the first order nature of the transition. In presence of disorder, there can be coexisting parent and product phases across the FOPT~\cite{sbroy}. CaMn$_7$O$_{12}$ contains both Mn$^{3+}$ and Mn$^{4+}$, and near a phase boundary one could have a chance for the formation of isolated local magnetic clusters through double exchange mechanism.

\par     
La doping at the Ca site was performed to change the Mn$^{3+}$ and Mn$^{4+}$ ratio of the sample, and it is likely that the La$^{3+}$ would enhance the percentage of Mn$^{3+}$ ions in the sample.  Our XPS investigation  indicates higher Mn$^{3+}$ fraction in the $x$ = 0.04 sample. XPS is predominantly a surface sensitive measurement, and one should be cautious in predicting the quantitative value of  Mn$^{3+}$/Mn$^{4+}$ ratio in the bulk from the peak area of the XPS data. However, our investigation based on the analysis of several independent aspects of XPS data can at least provide a supportive evidence for the enhanced Mn$^{3+}$ concentration in the La-doped samples. Notably  Mn$^{3+}$/Mn$^{4+}$ ratio calculated from the effective paramagnetic moment indicates the systematic variation of Mn-valency with La concentration and it is in line with the XPS analysis. It is clear from the $M(T)$ data that La doping at the Ca site causes systematic decrement of $M$ below $T_2$, although $M$ remains almost same above $T_2$ with changing $x$. Such observation supports the proposed scenario of superparamagnetic like clusters below $T_2$. La-doping will reduce the Mn$^{4+}$ fraction responsible for double exchange and as a result the number of such superparamagnetic clusters will shrink.

\par
Contrary to the value of $M$, $P$ is found to be larger in the La doped samples. We find that the small La (8\%) doping can enhance $P$ by a factor 3. This provides a easy path to enhance the functionality of this  material, and consequently  important from  the  point of view of applications. It is expected that the ME effect in CaMn$_7$O$_{12}$ is connected to the low-$T$ spiral magnetic order, which is also present in several other multiferroics~\cite{tokura}. However, the nature of the electric polarization is anomalous in CaMn$_7$O$_{12}$, where ${\vec{P}}$ lies perpendicular to the plane of the spiral magnetic structure with exceptionally large magnitude. In addition, due to the existence of proper screw type spiral magnetic structure in CaMn$_7$O$_{12}$~\cite{ws1}, a simple spin current model can not alone describe the origin of multiferroicity~\cite{arima, tokura}.  There are couple of different models proposed for large $P$ in CaMn$_7$O$_{12}$, namely (i) ferroaxial coupling of the magnetic chirality to the structural rotation~\cite{rdj}, (ii) combination of exchange-striction and DM interaction~\cite{lu}. At this point it is really difficult to comment on the mechanism behind the enhanced $P$ on La substitution at the Ca site. In the exchange-striction model based on DFT calculations, it is found that the induced $P$ depends on the value of  $\sin \alpha$, where $\alpha$ is an angle between Mn$^{4+}$ spins with the $x$ axis of a predefined coordinate system~\cite{lu}. The optimum value of $\alpha$ is found to be close to  30$^{\circ}$, which is obtained by  minimizing the energy Hamiltonian comprising spin exchange interaction and DM interaction. It is likely that due to the replacement of Mn$^{4+}$ by Mn$^{3+}$, the optimum value of $\alpha$ would change, which may be responsible for the drastic enhancement in $P$.

\par
It is to be noted that even a proper screw type magnetic structure can induce ferroelectricity through the variation in the metal-ligand hybridization in presence of spin-orbit coupling in relatively low-symmetry crystals~\cite{arima,jc}. This has been found to be quite useful in describing magnetic order induced polarization in materials such as CuFeO$_2$. Such scenario can not be ruled out for the observed multiferroicity in CaMn$_7$O$_{12}$. The metal-ligand ($p-d$) hybridization may depend upon the spin state of Mn and hence it can contribute to the large enhancement of $P$ in electron doped samples of CaMn$_7$O$_{12}$.  

\par
In conclusion, we  studied the magnetic and electric properties of CaMn$_7$O$_{12}$ and some of its La doped derivatives.    
We find that the low temperature magnetic transition is first order in nature and the ground state is magnetically metastable. A probable scenario of the formation of superparamagnetic like spin clusters is proposed. La doping reduces the superparamagnetic contribution at low temperature, however it enhances the  electric polarization to large extend. Apparently, the enhanced Mn$^{3+}$ ions in the sample at least within the doping range of present investigation is somehow related to the  enhanced polarization. This piece of information may be important for further  understanding of  the origin of anomalous multiferroicity in CaMn$_7$O$_{12}$. Lastly, the enhanced $P$ on La doping may be important for practical applications as far as the functionality of the materials is concerned.

\section{Acknowledgment}
 Unit of Nanoscience at IACS is duly acknowledged for magnetic and XPS measurements. We are thankful to the Surface Physics Division, SINP, Kolkata for EDS measurement. We also thank CSIR, India for financial support (grant number: 03(1209)/12/EMR-II). SC wishes to thank CSIR, India for his research fellowship.


\section*{References}
\begin{thebibliography}{99}

\bibitem{mathur} Eerenstein W , Mathur N D  and Scott J F 2006  {\it Nature} {\bf 442} 759

\bibitem{cheong} Cheong S -W  and  Mostovoy M 2007 {\it Nat. Mater.} {\bf 6} 13 

\bibitem{fiebig} Fiebig M 2005 {\it J. Phys. D: Appl. Phys.} {\bf 38} 123

\bibitem{lawes1} Lawes G and Srinivasan G 2011 {\it J. Phys. D: Appl. Phys.} {\bf 44} 243001

\bibitem{catalan} Catalan G and Scott J F 2009 {\it Adv. Mater.} {\bf 21} 2463

\bibitem{kimura} Kimura T, Kawamoto S, Yamada I, Azuma M, Takano M and Tokura Y 2003 {\it Phys. Rev. B} {\bf 67} 180401

\bibitem{khomskii} Khomskii D I 2006 {\it J. Mag. Mag. Mater.} {\bf 306} 1 

\bibitem{goto} Goto T, Kimura T, Lawes G, Ramirez A P and Tokura1 Y 2004 {\it Phys. rev. Lett.} {\bf 92} 257201

\bibitem{hur} Hur N, Park S, Sharma P A, Ahn J S, Guha S and  Cheong S-W 2004 {\it Nature} {\bf 429} 392

\bibitem{lawes2} Lawes G, Harris A B, Kimura T, Rogado N, Cava R J, Aharony A, Entin-Wohlman O, Yildirim T, Kenzelmann M, Broholm C and Ramirez A P 2005 {\it Phys. Rev. Lett.} {\bf 95} 087205

\bibitem{zhang} Zhang G, Dong S, Yan Z, Guo Y, Zhang Q, Yunoki S, Dagotto E and Liu J M 2011 {\it Phys. rev. B} {\bf 84} 174413
 
\bibitem{rdj} Johnson R D, Chapon L C, Khalyavin D D, Manuel P, Radaelli P G and Martin C 2012 {\it Phys. Rev. Lett.} {\bf 108} 067201 

\bibitem{bochu} Bochu B, Buevoz J L, Chenavas J, Collomb A, Joubert J C and Marezio M 1980 {\it Solid State Communications} {\bf 36} l33

\bibitem{ws} \L awi\'nski W S, Przenios\l o R, Sosnowska I, Bieringer M, Margiolaki I, Fitch A N and Suard E 2008 {\it J. Phys.:
Condens. Matter} {\bf 20} 104239

\bibitem{rp1} Przenios\l o R, Sosnowska I, Hohlwein D, Hau\ss T and Troyanchuk I O 1999 {\it Solid State Communications} {\bf 111} 687

\bibitem{rp2} Przenios\l o R, Sosnowska I, Suard E, Hewat A, Fitch A N 2004 {\it Physica B} {\bf 344} 358

\bibitem{ms} S\'anchez-And\'ujar M, Y\'a\~nez-Vilar S, Biskup N, Castro-Garc\'ia S, Mira J, Rivas J and Se\~nar\'is-Rodr\'iguez M A 2009 {\it J. Mag. Mag. Mater}  {\bf 321} 1739

\bibitem{lu} Lu X Z, Wangbo M -H, Dong S, Gong X Z G and Xiang H J 2012 {\it Phys. Rev. Lett.} {\bf 108} 187204


\bibitem{taguchi} Taguchi H, Nago M, Shimada M, Takeda Y and Yamamoto O 1988 {\it J. Solid State Chemistry} {\bf 77} 336

\bibitem{decorse} Decorse P, Caboche G and Dufour L -C 1999 {\it Solid State Ionics} {\bf 117} 161

\bibitem{vrg} Galakhov V R, Demeter M, Bartkowski S, Neumann M, Ovechkina M A, Kurmaev E Z, Lobachevskaya N I, Mukowskii Y M, Mitchell J and Ederer D L 2002 {\it Phys. Rev. B} {\bf 65} 113102

\bibitem{beyreuther} Beyreuther E, Grafstr\"o m S, Eng L M, Thiele C and D\"o rr K 2006 {\it Phys. Rev. B} {\bf 73} 155425

\bibitem{park} Park K H and Oh S -J 1993 {\it Phys. Rev. B} {\bf 48} 14833  

\bibitem{chaddah} Chaddah P, Kumar K and Banerjee A 2008 {\it Phys. Rev. B} {\bf 77} 100402

\bibitem{roy} Roy S B, Chattopadhyay M K, Chaddah P, Moore J D, Perkins G K, Cohen L F, Gschneidner, Jr. K A and Pecharsky V K 2006 {\it Phys. Rev. B} {\bf 74}    012403

\bibitem{chatterjee} Chatterjee S,  Giri S, Majumdar S and De S K 2008 {\it Phys. Rev. B} {\bf 77} 224440

\bibitem{ito} Ito A, Aruga H, Torikai E, Kikuchi M, Syono Y and Takei H 1986 {\it Phys. Rev. Lett.} {\bf 57} 483  

\bibitem{rvc} Chamberlin R V, Mozurkewich G and Orbach R 1984 {\it Phys. Rev. Lett.} {\bf 52} 867 

\bibitem{phillips} Phillips J C 1996 {\it Rep. Prog. Phys.} {\bf 59} 1133

\bibitem{xd} Du X, Li G, Andrei E Y, Greenblatt M  and Shuk P 2007 {\it Nat. Phys.} {\bf 3} 111

\bibitem{salamon} Sun Y, Salamon M B, Garnier K and Averback R S 2003 {\it Phys. Rev. Lett.} {\bf 91} 167206

\bibitem{gdcu} Bhattacharyya A, Giri S and Majumdar S 2011 {\it Phys. Rev. B} {\it 83} 134427

\bibitem{sasaki} Sasaki M, J\"onsson P E, Takayama H and Mamiya H 2005 {\it Phys. Rev. B} {\bf 71} 104405

\bibitem{ban} Bandyopadhyay M and Dattagupta S 2006 {\it Phys. Rev. B} {\bf 74} 214410

\bibitem{kohn} Inomata A and Kohn K 1996 {\it J. Phys.: Condens. Matter} {\bf 8} 2673

\bibitem{kk} Kitamura K, Hatano H, Takekawa S, Sch\"utze D and Aono M 2010 {\it Appl. Phys. Lett.} {\bf 97} 082903

\bibitem{bedanta} Bedanta S  and Kleemann W 2009  {\it J. Phys. D: Appl. Phys.} {\bf 42} 013001

\bibitem{banerjee} Bajpai A and Banerjee A 2000 {\it Phys. Rev. B} {\bf 62} 8996

\bibitem{jia} Jia L, Liu G J, Wang J Z, Sun J R, Zhang H W and Shen B G 2006 {\it Appl. Phys. Lett.} {\bf 89} 122515

\bibitem{granado} Granado E, Ling C D, Neumeier J J, Lynn J W and Argyriou D N 2003 {\it Phys. Rev. B} {\bf 68} 134440

\bibitem{qin} Qin Y, Tyson T A, Pranzas K and Eckerlebe H 2008 {\it J. Phys.: Condens. Matter} {\bf 20} 195209

\bibitem{dagotto} Dagotto E, Hotta T and Moreo A 2001 {\it Phys. Rep.} {\bf 344} 1

\bibitem{sbroy} Roy S B and Chaddah P 2004 {\it Phase Transitions} {\bf 77} 767

\bibitem{tokura} Tokura Y and Seki S 2010 {\it Adv. Mater.} {\bf 22} 1554

\bibitem{ws1} \L awi\'nski W , Przenios\l o R, Sosnowska I and Pet\v{r}\'{i}\v{c}ek V 2012 {\it Acta Cryst. B} {\bf 68} 240

\bibitem{arima} Arima T 2007 {\it J. Phys. Soc. Jpn.} {\bf 76} 073702

\bibitem{jc} Jia C, Onoda S, Nagaosa N and Han J H 2007 {\it Phys. Rev. B} {\bf 76} 14424

\end{thebibliography}
\end{document}